\documentclass[letterpaper,twocolumn,10pt]{article}

\usepackage{ifthen}
\usepackage[normalem]{ulem} 
\usepackage{xcolor}
\usepackage{amssymb}

\newboolean{showedits}
\setboolean{showedits}{true} 
\ifthenelse{\boolean{showedits}}
{
	\newcommand{\del}[1]{\textcolor{red}{\sout{#1}}} 
}{
	\newcommand{\del}[1]{} 
	
}

\newboolean{showcomments}
\setboolean{showcomments}{true}
\newcommand{\id}[1]{$-$Id: scgPaper.tex 32478 2010-04-29 09:11:32Z oscar $-$}

\ifthenelse{\boolean{showcomments}}
{\newcommand{\nbc}[3]{
 {\colorbox{#3}{\bfseries\sffamily\scriptsize\textcolor{white}{#1}}}
 {\textcolor{#3}{\sf\small$\blacktriangleright$\textit{#2}$\blacktriangleleft$}}}
 }
{\newcommand{\nbc}[3]{}
 \renewcommand{\del}[1]{} 
 }

\definecolor{ibcolor}{rgb}{0.9,0.5,0}
\definecolor{cfcolor}{rgb}{0,0.5,0.9}

\definecolor{tdcolor}{rgb}{1.0,0,0}

\usepackage{usenix2019,epsfig,endnotes}
\usepackage{amsmath}
\usepackage{graphicx}
\usepackage{enumitem}
\usepackage{float}
\usepackage{hyperref}
\usepackage[]{algorithm2e}
\usepackage[page]{appendix}

\begin{document}

\setlength{\pdfpageheight}{\paperheight}
\setlength{\pdfpagewidth}{\paperwidth}

\title{\Large \bf Dancing in the Dark: Private Multi-Party Machine Learning in an Untrusted Setting}

\author{
{\rm Clement Fung} \\
University of British Columbia \\
clement.fung@alumni.ubc.ca
\and
{\rm Jamie Koerner\thanks{Work done while at University of British Columbia}} \\
University of Toronto \\
jamie.koerner@utoronto.ca
\and
{\rm Stewart Grant\footnotemark[1]} \\
University of California, San Diego \\
ssgrant@eng.ucsd.edu
\and
{\rm Ivan Beschastnikh} \\
University of British Columbia \\
bestchai@cs.ubc.ca
}





\maketitle

\subsection*{Abstract}

Distributed machine learning (ML) systems today use an unsophisticated
threat model: data sources must trust a central ML process. We propose
a \emph{brokered learning} abstraction that allows data sources to
contribute towards a globally-shared model with provable privacy
guarantees in an untrusted setting. We realize this abstraction by
building on federated learning, the state of the art in multi-party ML,
to construct \emph{TorMentor}: an anonymous hidden service that
supports private multi-party ML. 

We define a new threat model by characterizing, developing and
evaluating new attacks in the brokered learning setting, along with
new defenses for these attacks. We show that TorMentor effectively
protects data providers against known ML attacks while providing them
with a tunable trade-off between model accuracy and privacy.
We evaluate TorMentor with local and geo-distributed deployments on
Azure/Tor. In an experiment with 200 clients and 14 MB of data per
client, our prototype trained a logistic regression model using
stochastic gradient descent in 65s.

\section{Introduction}
\label{sec:intro}

Machine learning (ML) models rely on large volumes of diverse,
representative data for training and validation. However, to build
multi-party models from user-generated data, users must provide and
share their potentially sensitive information. Existing 
solutions~\cite{Low:2012, Li:2014}, even when incorporating
privacy-preserving mechanisms~\cite{Abadi:2016}, assume
the presence of a trusted central entity that all users share
their data and/or model updates with. For example, the Gboard
service~\cite{Gboard:2017} uses sensitive data from Android keyboard
inputs to generate better text suggestions; users who wish to train
an accurate Gboard suggestion model must send their mobile keyboard
data to Google.

Federated learning~\cite{McMahan:2017} keeps data on the client device
and trains ML models by only transferring model parameters to a trusted
central coordinator. In the quest for training the optimal ML model as
quickly as possible, the coordinator is not incentivized to provide
privacy for data providers: data is collected and is only kept on the
device as a performance optimization~\cite{McMahan:2017}.  

Furthermore, federated learning is not private or secure against
adversaries, who can pose as honest data providers or an honest
coordinator and who use auxiliary information from the learning process to
infer information about the training data of other 
users~\cite{Hitaj:2017}, despite data never being
explicitly shared. One may consider obfuscating data labels before
learning, but this is also insufficient to guarantee 
privacy~\cite{Calandrino:2011}. General privacy-preserving computation
models exist, but they rely on substantial amounts of additional
infrastructure. These include homomorphically encrypted secure
aggregation~\cite{Bonawitz:2017}, secure multi-party 
computation~\cite{Mohassel:2017}, or trusted SGX
infrastructure~\cite{Ohrimenko:2016}, all of which are infeasible for
individuals and small organizations to deploy.

Today there is no accessible solution to collaborative multi-party
machine learning that maintains privacy and anonymity in an untrusted
setting. In developing this solution, we propose a novel setting
called \textit{brokered learning} that decouples the role of
coordination from the role of defining the learning process. We
introduce a short-lived, honest-but-curious \emph{broker} that mediates
interactions between a \emph{curator}, who defines the shared learning
task, and \emph{clients}, who provide training data. In decoupling
these roles, curators are no longer able to directly link clients to
their model updates, and cannot manipulate the learning
process.

Clients and curators never directly communicate: they are protected
from each other by a broker that is only used to coordinate the
learning task. The broker is administered by an honest-but-curious
neutral party, meaning that it does not initiate actions and
follows the prescribed learning process. The broker detects and
rejects anomalous behavior and terminates when the learning process as
instructed, but is curious and will examine client data if able. Our
system design supports privacy and anonymity by building on accessible
public infrastructure to minimize the cost of setting up and
maintaining a broker.

We realize the brokered learning setting by designing, implementing,
and evaluating \emph{TorMentor}, a system which creates \emph{brokers}
to interface with the curator and the clients
in a brokered learning process. TorMentor is
implemented as a hidden Tor service, but can use any general-purpose
anonymous communication service to safeguard the identities of
curators and clients. 


Although the model curator is removed from the learning process, a
myriad of other attacks are still possible. We adapt known ML attacks
to brokered learning and build on several state of the art techniques
to thwart a variety of these attacks when they are mounted by clients,
brokers and curators. Client-side differential
privacy~\cite{Dwork:2014, Geyer:2017} protects users from inversion
attacks~\cite{Fredrikson:2014, Fredrikson:2015}, reject on negative
influence (RONI)~\cite{Barreno:2010} and monitored client
statistics~\cite{Mozaffari-Kermani:2015} prevent model poisoning
attacks~\cite{Biggio:2012, Huang:2011} and proof of
work~\cite{Back:2002} mitigates sybil attacks~\cite{Douceur:2002}.

Our evaluation of TorMentor demonstrates that these defenses protect
clients and curators from each other. For example, in one experiment
with 50\% malicious poisoning clients, a TorMentor broker was able to
converge to an optimum after mitigating and recovering from 
malicious behavior through our novel adaptive proof of work mechanism.
We also evaluated the performance of our prototype in a geo-distributed
setting: across 200 geo-distributed clients with 14 MB of data
per client, the training process in TorMentor takes 67s. By
comparison the training on a similar federated learning system without
Tor would take 13s. The observed overhead of TorMentor ranges from
5-10x, and depending on the level of privacy and security required,
TorMentor's modular design allows users to further tune the system to
meet their expected needs on the security-performance trade-off.

In summary, we make four contributions:
\begin{itemize}[label=$\star$]
 \item We develop a \textit{brokered learning} setting for
    privacy-preserving anonymous multi-party machine learning in an
    untrusted setting. We define the responsibilities, interactions,
    and threat models for the three actors in brokered learning:
    curators, clients, and the broker.

 \item We realize the brokered learning model in the design and
    implementation of \textit{TorMentor} and evaluate TorMentor's
    training and classification performance on a public dataset.

 \item We translate known attacks on centralized ML 
    (poisoning~\cite{Huang:2011, Nelson:2008} and 
    inversion~\cite{Fredrikson:2014, Fredrikson:2015}) and known
    defenses in centralized ML (RONI~\cite{Barreno:2010}, differential
    privacy~\cite{Dwork:2014}) to the brokered learning setting. We
    evaluate the privacy and utility implications of these attacks and
    defenses.

 \item We design, implement, and evaluate three new defense mechanisms
    for the brokered learning setting: distributed RONI,
    adaptive proof of work, and thresholding the number of clients.

\end{itemize}

\section{Background}
\label{sec:background}

\noindent \textbf{Machine Learning (ML).} 
Many ML problems can be represented as the minimization of a loss
function in a large Euclidean space. For example, a binary
classification task in ML involves using features from data examples to
predict discrete binary outputs; a higher loss results in more
prediction errors. Given a set of training data and a proposed model, ML
algorithms \emph{train}, or iteratively find an optimal set of
parameters, for the given training set. One approach is to use
stochastic gradient descent (SGD)~\cite{Bottou:2010}, an iterative
algorithm which samples a batch of training examples of size $b$, uses
them to compute gradients on the parameters of the current model, and
takes gradient steps in the corresponding gradient direction. The
algorithm then updates the model parameters and another iteration is
performed (Appendix A contains all background formalisms).

Our work uses SGD as the training method. SGD is a
general learning algorithm that is used to train a variety
of models, including deep learning~\cite{Song:2013}.\\

\noindent \textbf{Distributed multi-party ML.}
To support larger models and datasets, ML training has been
parallelized using a parameter server architecture~\cite{Li:2014}: the
global model parameters are partitioned and stored on multiple parameter
server machines. At each iteration, client machines pull the 
parameters from server machines, compute and apply one or more
iterations, and push their updates back to the server. This can be
done with a synchronous or asynchronous protocol~\cite{Hsieh:2017,
Recht:2011}, both of which are supported in our work.


\textit{Federated Learning}~\cite{McMahan:2017}. The partitioning of
training data enables multi-party machine learning: data is
distributed across multiple data owners and cannot be shared. Federated
learning supports this setting through a protocol in which clients send
gradient updates in a distributed SGD algorithm. These updates are
collected and averaged by a central server, enabling training over
partitioned data sources from different owners. \\

\noindent \textbf{Attacks on ML.} 
%
Our work operates under a unique and novel set of assumptions when
performing ML and requires a new threat model. Despite this novel
setting, the attacks are functionally analogous to state of the art ML
attacks known today. 

\emph{Poisoning attack.}  In a poisoning attack~\cite{Biggio:2012,
Mozaffari-Kermani:2015}, an adversary meticulously creates adversarial
training examples and inserts them into the training data set of a 
target model. This may be done to degrade the accuracy of the final
model (a random attack), or to increase/decrease the probability of a
targeted example being predicted as a target class (a targeted 
attack)~\cite{Huang:2011}. For example, such an attack could be
mounted to avoid anomaly detectors~\cite{Rubinstein:2009} or to evade
email spam filtering~\cite{Nelson:2008}.  

In federated learning, clients possess a disjoint set of the
total training data; they have full control over this set, and can
therefore perform poisoning attacks with minimal difficulty.

\emph{Information leakage.}  In an information leakage attack, such as
model inversion, an adversary attempts to recover training examples
from a trained model $f(x)$. Given a targeted class $y$, one inversion
technique involves testing all possible values in a feasible feature
region, given that some information about the targeted example is
known. The attack then finds the most probable values of the targeted
example features~\cite{Fredrikson:2014}.

Another model inversion attack uses prediction confidence scores from
$f(x)$ when predicting $y$ to perform gradient descent to train a model
$\hat{f(x)}$ that closely resembles $f(x)$. The victim's features are
probabilistically determined by using the model in prediction. The
resulting vector $x$ that comes from this process is one that
most closely resembles an original victim training 
example~\cite{Fredrikson:2015}. 

Information leakage attacks have been extended to federated learning:
instead of querying information from a fully trained model, an
adversary observes changes in the shared model during the training
process itself~\cite{Hitaj:2017}. In doing so, the adversary can
reconstruct training examples that belong to other clients. \\

\noindent \textbf{Defenses for ML.}  
\emph{A RONI (Reject on Negative Influence)} 
defense~\cite{Barreno:2010} counters ML poisoning attacks. Given a set
of untrusted training examples $D_{un}$, this defense trains two
models, one model using all of the trusted training data $D$, and
another model using the union dataset $D' = D \cup D_{un}$ which
includes the untrusted data. If the performance of $D'$ is
significantly worse than the performance of $D$ on a validation
dataset, the data $D_{un}$ is flagged as malicious and rejected.
However, this defense relies on access to the centralized dataset,
which is infeasible in the federated learning setting.

\emph{Differential privacy}~\cite{Dwork:2014} is a privacy guarantee
that ensures that, for a given dataset, when used to answer questions
about a given population, that no adversary can identify individuals
that are members of the dataset. 
Differential privacy is user-centric: the violation of a single user's
privacy is considered a privacy breach. Differential privacy is
parameterized by $\varepsilon$, which controls the privacy-utility
trade-off. In the ML context, utility equates to model 
performance~\cite{Song:2013}. When $\varepsilon$ approaches 0, full
privacy is ensured, but an inaccurate model is produced. When
$\varepsilon$ approaches infinity, the optimal model is produced
without any privacy guarantees.

Differential privacy has been applied to several classes of ML
algorithms~\cite{Shokri:2015, Bellet:2017} in decentralized settings to
theoretically guarantee that a client's privacy is not compromised when
their data is used to train a model. This guarantee extends to both the
training of the model and to usage of the model for predictions.

Differentially private SGD~\cite{Song:2013, Geyer:2017} is a method
that applies differential privacy in performing SGD. Before sending
gradient updates at each iteration, clients perturb their gradient
values with additive noise, which protects the privacy of the input
dataset. The choice of batch size impacts the effect of the privacy
parameter on the model performance. Our work uses differentially
private SGD to provide the flexible privacy levels against attacks in
our setting.\\

\noindent \textbf{Anonymous communication systems.}
Clients are still at privacy risk when sending model updates directly
to an untrusted broker, so we add an additional layer of indirection in
our work. Onion routing protocols are a modern method for providing
anonymity in a distributed P2P setting. When communicating with a
target node, clients route traffic through a randomly selected set of
relay nodes in the network, which obfuscates the source and destination
nodes for each message.

Our work supports the use of any modern implementation for providing
anonymous communication. We select the Tor 
network~\cite{Dingledine:2004} as the system in our implementation. \\

\noindent \textbf{Sybil attacks and proof of work.}
An anonymous system that allows users to join and leave may be attacked
by sybils~\cite{Douceur:2002}, in which an adversary joins a system
under multiple colluding aliases. One approach to mitigate sybil
attacks is to use proof of work~\cite{Back:2002}, in which a user must
solve a computationally expensive problem (that is easy to verify) to
contribute to the system. This mechanism provides the guarantee that if
at least 50\% of the total computation power in the system is
controlled by honest nodes, the system is resilient to sybils.


\begin{figure}[t]
  \centering
  \includegraphics[width=.9\linewidth]{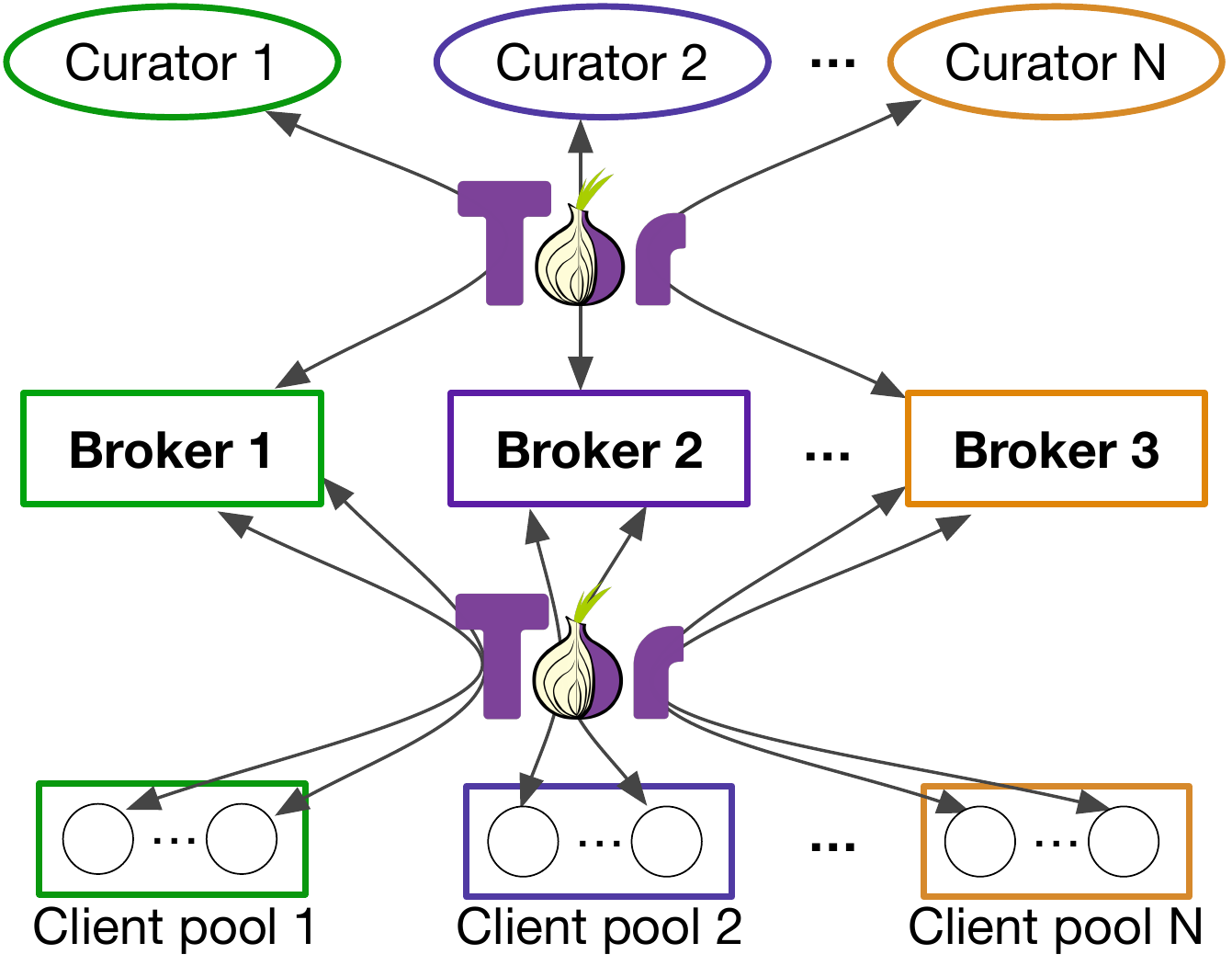}
  \caption{Brokered learning and TorMentor design. Brokers
    (implemented as hidden Tor services) mediate between curators
    (top) and sets of clients (bottom). }
  \label{fig:sysdiagram}
\end{figure}

\section{Brokered Learning}
\label{sec:setting}

In this work, we define \textit{brokered learning}, which builds on the
federated learning setting~\cite{McMahan:2017}, but assumes no trust
between clients and curators.

\subsection{Decoupling federated learning}

In federated learning, a central organization (such as Google) acts as
the parameter server and performs two logically separate tasks:
\textbf{(1)} define the data schema and the learning objective, 
\textbf{(2)} coordinate distributed ML. The federated learning server
performs both tasks at a central service, however, there are good
reasons to separate them.

Fundamentally, the goals of data privacy and model accuracy are at
tension. Coordinating the ML training process in a private and secure
manner compromises the model's ability to learn as much as possible
from the training data. In current learning settings, the coordinator
is put in a position to provide privacy, yet they are not incentivized
to do so.

To take things even further, a malicious curator can observe the
contributions of any individual client, creating an opportunity to
perform information leakage~\cite{Hitaj:2017} attacks on clients, such
as model inversion~\cite{Fredrikson:2014, Fredrikson:2015} or
membership inference~\cite{Shokri:2017}. These attacks can be
mitigated in federated learning with a secure aggregation
protocol~\cite{Bonawitz:2017}, but this solution does not handle
poisoning attacks and requires several coordinated communication
rounds between clients for each iteration.

Client anonymity may also be desirable when privacy preferences are
shared. For example, if attempting to train a model that uses past
criminal activity as a feature, one user with strong privacy
preferences in a large group of users with weak privacy preferences
will appear suspicious, even if their data is not
revealed. 


A key observation in our work is that \emph{because data providers and model
curators agree on a learning objective before performing federated
learning, there is no need for the curator to also coordinate the
learning.}

\emph{Brokered learning} decouples the two tasks into two distinct
roles and includes mechanisms for anonymity to protect data providers
from the curator while orchestrating federated learning. In this
setting the users in the system (model curators and data providers) do
not communicate with one another, which facilitates a minimal trust
model, strengthening user-level privacy and anonymity. As well, users
do not reveal their personal privacy parameters to the broker, since 
\emph{all privacy-preserving computation is performed on the client}.

At the same time brokered learning maintains the existing privacy
features of federated learning: data providers do not need to
coordinate with each other (or even be aware of each other).
This is important to counter malicious clients who attack
other clients~\cite{Hitaj:2017}.

\subsection{Defining brokered learning}

\emph{Brokered learning} builds on federated 
learning~\cite{McMahan:2017} but provides additional privacy
guarantees. This model introduces a \textbf{broker} to mediate the
learning process. 

\textbf{Curators} define the machine learning model. A curator has a
learning task in mind, but lacks the sufficient volume or variety of
data to train a quality model. Curators would like to collaborate with
clients to perform the learning task and may want to remain
anonymous. We provide curators with anonymity in TorMentor by
deploying a broker as a Tor hidden service, and by using the broker
as a point of indirection (Figure~\ref{fig:sysdiagram}).


Curators may know the identities of clients that wish to contribute to
the model or may be unaware of the clients that match their learning
objectives. Brokered learning supports these and other use cases, For
example, curators may know some subset of the clients, or set a
restriction on the maximum number of anonymous clients who can
contribute\footnote{We assume that the broker identifies or advertises
the learning process to clients out of band, externally to TorMentor.}.

\textbf{Clients} contribute their data to the machine learning task
and specify the criteria for their participation. Instead of fully
trusting the curator as they would in federated learning, clients
communicate with an honest-but-curious broker. The broker is trusted
with coordinating the learning process and does not initiate actions.
However, the broker is curious and will examine client data and
indentities if possible. This threat model is similar to what was used
in the secure aggregation protocol for federated 
learning~\cite{Bonawitz:2017}.

Brokered learning allows these clients to jointly contribute to a
shared global model, without being aware of nor trusting each other.
Each client only needs to be concerned about its personal privacy
parameters. Some clients may be more concerned with privacy than
others; brokered learning supports differentially private
machine learning with heterogeneous privacy levels, which has been
shown to be feasible~\cite{Geyer:2017}. 


\textbf{A broker} is a short-lived process that coordinates the
training of a multi-party ML model. For each model defined by a
curator in TorMentor, a single broker is created and deployed as a
hidden service in an anonymous network \footnote{In this paper we do not
  consider who is running the broker; but we do assume that it is
  an honest-but-curious third party that is distinct from the
  curator and the participating clients.}.
Clients perform actions such as requesting access to the system, defining
client-specific privacy parameters and sending model updates for
distributed SGD. We define a precise client API in 
Section~\ref{sec:design}. When model training is complete, the broker
publishes the model and terminates. In our vision, brokers are not
intended to be long lasting, and their sole function should be to
broker the agreement between users to facilitate anonymous multi-party
ML. Brokers may even explicity be managed by governments or as part of
a privacy enhancing business model, both of whom are incentivized to
provide privacy, anonymity and fairness in distributed ML.


\subsection{Example use cases}

\noindent \textbf{Medical Sharing Network}. Hospitals store substantial
patient medical data. However, due to strict regulations, they 
cannot share this data with each other. No individual
hospital wishes to host the infrastructure for model coordination, and
no individual hospital is trusted to securely coordinate the analysis.
An alternative solution is for the network of hospitals to collaborate in
a brokered learning system. For this the hospitals would define the
learning task, one hospital would agree to deploy the broker as a
hidden service, and all other willing hospitals would join and
contribute model updates, training a shared model. \\

\noindent \textbf{Internet of Things}. With the growth of the Internet
of Things (IoT), and a largely heterogeneous set of device providers,
there is currently no solution for privacy-preserving multi-device ML,
hosted by a neutral provider. Without anonymous multi-party ML, each
system device provider would need to host their own ML coordinators and
would have no mechanism for sharing models across other providers. 

Brokered learning allows these devices to collaborate on model training
without explicitly trusting each other. Devices reap the benefits of
shared trained models, without risking data privacy loss. The broker
can be run by any single company, or a neutral trusted third party,
neither of which have power to compromise device-level privacy.

\section{Threat model, guarantees, assumptions}
\label{sec:threat}

\begin{figure}[t]
  \centering
  \includegraphics[width=.9\linewidth]{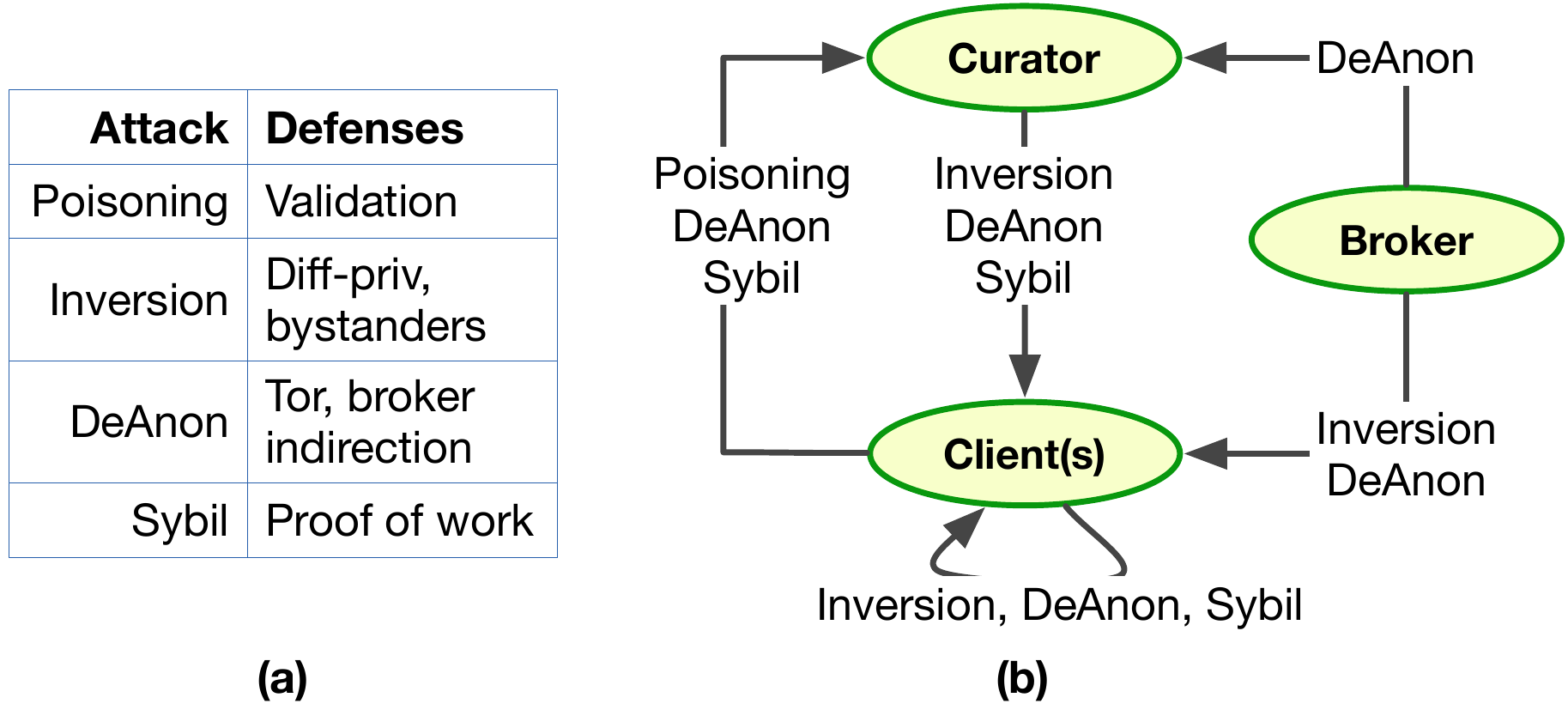}
  \caption{(a) Attacks/defenses in TorMentor. (b) Threat model: an
    edge is an attack by source against target(s).}
  \label{fig:threatmodel}
\end{figure}


We realized brokered learning in a system called \emph{TorMentor},
which uses differentially private distributed SGD~\cite{Song:2013} to
train a model in an anonymous multi-party setting. We select 
Tor~\cite{Dingledine:2004} as the anonymous communication network for
TorMentor. TorMentor is designed to counter malicious curators and
malicious clients who may attempt to gain additional information 
(information leakage) about others or negatively influence the
learning process (poisoning). The honest-but-curious broker coordinates
the learning process, but is \emph{not} trusted with the identity nor
the data of users.

Clients and curators do not attack the broker itself, rather they aim
to attack other curators, other clients, or the outcome of the learning
process. Brokers are also untrusted in the system: the client and
curator APIs protect them from potential broker attacks. 
Figure~\ref{fig:threatmodel} overviews TorMentor's threat model with
attacks/defenses and who can attack who and how. \\


\noindent \textbf{Deanonymization.} For anonymous communication to and from the
broker we assume a threat model similar to Tor~\cite{Dingledine:2004}:
an adversary has the ability to observe and control some, but
not all of the network. Adversaries may attempt to observe Tor traffic
as a client or as a broker in the network~\cite{Murdoch:2005,
Evans:2009, Johnson:2013}. Adversaries can also influence traffic
within Tor through their own onion router nodes, which do not
necessarily need to be active TorMentor clients.  \\

\noindent \textbf{Poisoning attacks.} In our threat model, poisoning attacks are
performed by clients against shared models. After a curator defines a
model, adversarial clients can use the defined learning objective to
craft adversarial samples that oppose the objective. We assume that
adversaries generate these adversarial samples using a common strategy
such as label flipping~\cite{Biggio:2012}, and join the training
process and poison the model by influencing its prediction
probabilities. \\

\noindent \textbf{Inversion attacks.} We assume that adversaries can target a
specific client victim who they know is using the system. Inversion
attacks can be executed from a variety of points: the adversary may be
administering the broker, the adversary may curate a model that the
victim joins, or the adversary joins model training as a client,
knowing that the victim has also joined. Although the broker does not
expose model confidence scores in the model prediction API, which are a
key piece of information for performing inversion attacks in a
black-box setting~\cite{Fredrikson:2015}, our threat model grants
adversaries white-box access to the model internals, which is more
powerful than a traditional model inversion attack. 

In TorMentor, adversarial clients have access to the global model and
can infer confidence scores or gradient step values by carefully
observing changes to the global model and reconstructing a copy of the
victim's local model. This attack is similar to a model stealing
attack~\cite{Tramer:2016}: after stealing an accurate approximation of
a victim's local model, an adversary can mount an inversion attack on
the approximation to reconstruct the victim's training examples. \\

\noindent \textbf{Sybil attacks.}  Since clients and curators join the system
anonymously, they can generate sybils, or multiple colluding virtual
clients, to attacks the system~\cite{Douceur:2002}. In federated
learning, all users are given an equal stake in the system, and
thus sybils make poisoning and inversion attacks linearly easier to
perform.

\subsection{Security guarantees} 

TorMentor guarantees that curator and client identities remain
anonymous to all parties in the system by using an anonymous
communication network. Our prototype uses Tor and provides the same
guarantees as Tor, however, it can use other anonymous
messaging systems~\cite{Vuvuzela:2015, Riposte:2015}. For example, since
Tor is susceptible to timing attacks, an adversary could target a
client by observing its network traffic to de-anonymize its
participation in the system. 

TorMentor exposes a small, restrictive API to limit a user's influence
on the system. TorMentor alleviates the risk of poisoning attacks and
inversion attacks by allowing clients to specify $k$, the minimum
number of clients required for training. When a client joins the
system, they specify the number of other clients required in the
system, which TorMentor guarantees will be honored. Clients also
locally use differential privacy to further protect their privacy.
Both parameters are defined by the client, guaranteeing that clients 
(and not the curator) controls this accuracy-privacy tradeoff.


TorMentor prevents sybils through proof of work, similar to the Bitcoin
protocol~\cite{Nakamoto:2009}. This mechanism makes it expensive,
though not impossible, to mount a sybil attack. Proof of work is
implemented at two points: proof of work as a prerequisite for system
admission, and proof of work as a prerequisite for contributing to the
model.

\subsection{Assumptions}

We assume that the only means to access information within the system
is through the APIs defined in Section~\ref{sec:design}. A TorMentor
instance and its corresponding brokers are exposed as a hidden service
with a public \textit{.onion} domain. We assume that this 
\textit{.onion} becomes a widely known and trusted domain\footnote{To
build further trust the TorMentor service can use an authoritatively
signed certificate.}.

We use proof of work to defend against sybils and therefore assume that
the adversary does not have access to more than half of the
computational power relative to the total computational power across
all users in an instance of the TorMentor training 
process~\cite{Back:2002}.

We make the same assumptions as Tor~\cite{Dingledine:2004}; for
example, we assume that an adversary does not control a large fraction
of nodes within the Tor network. 


\begin{table*}[t]
\begin{tabular}{ p{7cm} p{10cm} }
 \hline
 \textbf{API call}                                                 & 
 \textbf{Description}                                                                 \\
 \hline
 address $\leftarrow$ \textbf{curate}(mID, maxCli, minCli, validSet) & 
 Curate a new model. Curator provides modelID,  
 client count range, validation set. TorMentor returns a hidden
 service address for a newly specified broker. \\
 \hline\hline
 $P_{admit}$ $\leftarrow$ \textbf{join}(mID)                         
 &
 Client joins a curated model. Client provides modelID; TorMentor returns
 a SHA-256 admission hash puzzle $P_{admit}$.                          
 \\
 \hline
 conn, $M_t$ $\leftarrow$ \textbf{solve}(mID, $S_{admit}$, minCli, schema) & 
 Client finds the solution $S_{admit}$ to $P_{admit}$ and joins. Client
 provides modelID, solution to puzzle, min number of clients and its
 dataset schema; TorMentor returns a connection and global model state.
 \\
 \hline
 $M_{g,t+1}$, $P_{i,t+1}$ $\leftarrow$ \textbf{gradientUpdate}(mId, $S_
 {i,t}$, $\Delta_{i,t}$) & 
 Client pushes a local model update to the global model state. Client $i$
 provides modelID, solution to previous SHA-256 puzzle $S_{i,t}$ and
 gradient update $\Delta_{i,t}$ at iteration $t$; TorMentor returns new
 global model state $M_{g,t+1}$, and the next SHA-256 puzzle $P_
 {i,t+1}$.          
 \\
\end{tabular} 
\caption{TorMentor API. The curate call (top row) is the only curator
  API call. The bottom three calls are for clients.\label{tab:API} }
\end{table*}

\begin{figure}[t]
  \centering
  \includegraphics[width=.9\linewidth]{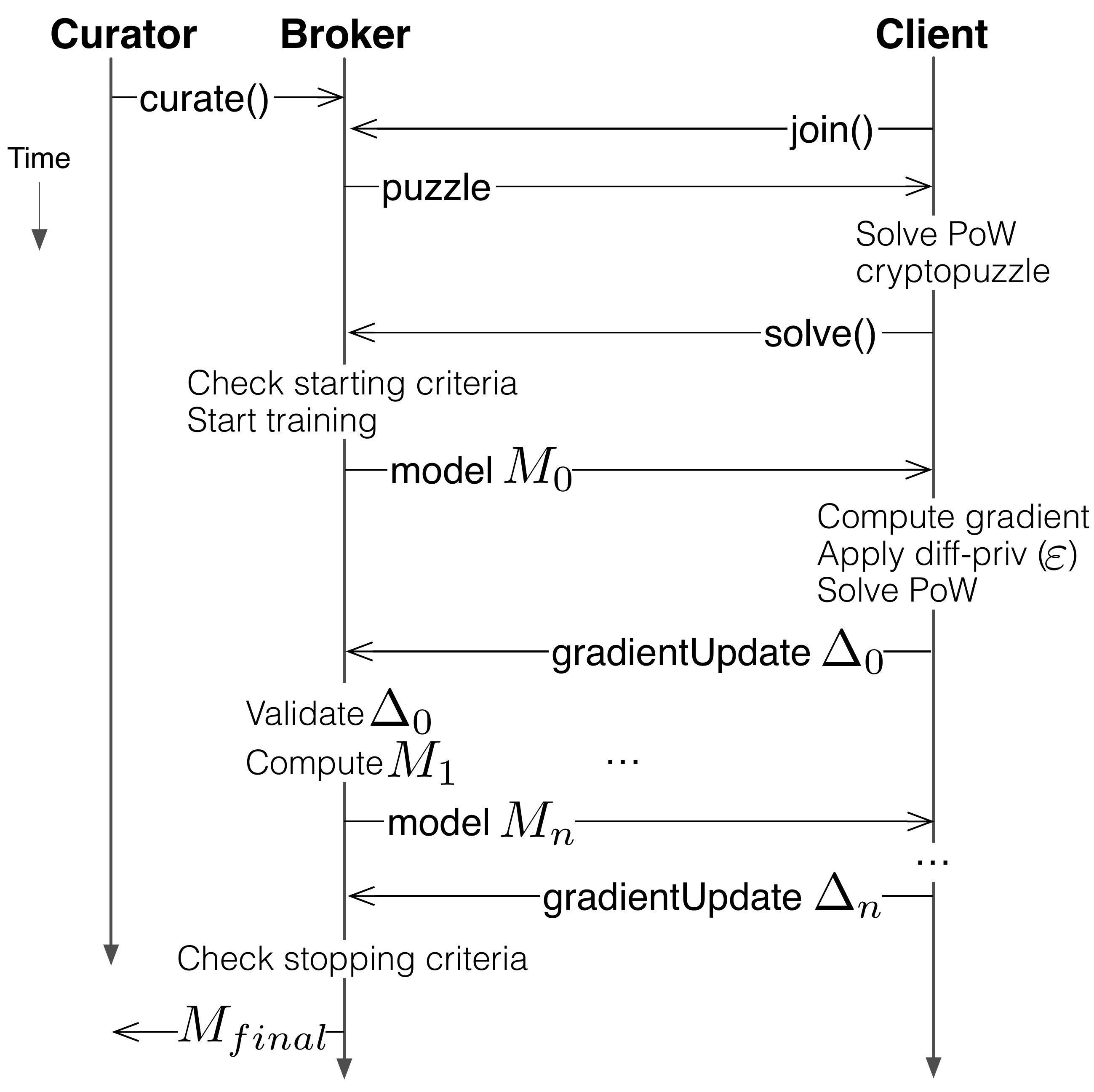}
  \caption{Overview of the TorMentor protocol between
    curator/broker/client.
    }
  \label{fig:protocol}
\end{figure}

\section{TorMentor Design}
\label{sec:design}

TorMentor design has three goals: (1) meet the defined learning
objective in a reasonable amount of time, (2) provide both anonymity
and data privacy guarantees to clients and curators, and (3) flexibly
support client-specific privacy requirements. \\

\noindent \textbf{Design overview.} The broker handles all
communication between clients and the curator, and acts as the
coordinator in an untrusted collaborative learning setting. Each
TorMentor broker is deployed as a Tor hidden service with a unique
and known \textit{.onion} domain. Several clients may join a model
once a curator defines it, with requirements for joining specified by
both the curator and the clients. Each broker and therefore each model
is associated with a pool of clients among whom the learning procedure
takes place (see Figure~\ref{fig:sysdiagram}).

Each broker runs a separate \emph{aggregator} process and 
\emph{validator} process. The aggregator serves the same purpose as a
parameter server~\cite{Li:2014}: storing and distributing the
parameters of the global model. The validator is a novel addition in
our work that observes and validates the values of gradient updates
sent by clients to the aggregator.

Next, we review the TorMentor API in Table~\ref{tab:API} and the
training process illustrated in Figure~\ref{fig:protocol}. We then
detail how TorMentor defends against adversarial clients and curators.

\subsection{Curator API}

The curator uses the \textbf{curate} call to bootstrap a new model by defining
a \emph{common learning objective}: the model type, the desired
training data schema and a validation dataset. These are critical to
perform ML successfully (even in a local setting). We therefore expect
that a curator can provide these.

Once the learning objective is defined, a Tor \textit{.onion} address
is established for the specified model, and the system waits for
clients to contact the hidden service with a message to join. The
validation dataset is used by the validator to reject adversaries, and
to ensure that the ML training is making progress towards model
convergence.

Too few clients may lead to a weak model with biased data, while a
large number of clients will increase communication overhead. The
curator can use the API to adjust an acceptable range for the number of
clients contributing to the model.

\subsection{Client API} 

A client uses the \textbf{join} call to join a curated model
\footnote{We assume that the client is able to learn the
curator-provided modelID out of band. This may be through a
third-party system or directly through another anonymous service.}
A client's data is validated against the objective when joining. Our
prototype only checks that the specified number of features matches
those of the client, but more advanced automatic schema validation
techniques~\cite{Rahm:2001} can be used.

The client uses the \textbf{solve} call to perform a proof-of-work
validation, similar to that of the blockchain 
protocol~\cite{Nakamoto:2009}, in which a cryptographic SHA-256
admission hash is inverted, the solution is verified to contain
a required number of trailing `0' digits, and a new puzzle is
published. Once the proof-of-work is completed, the client is accepted
as a contributor to the model. Once the desired number of clients have
been accepted to the model, collaborative model training is performed
through the TorMentor protocol: each client computes their SGD update
on the global model and pushes it to the parameter server through the 
\textbf{gradientUpdate} call.

Clients compute gradient updates \emph{locally}. Clients also
maintain a personal privacy level $\varepsilon$ and a personal batch
size $b$ to tune their differentially-private updates during model
training. With the privacy-utility tradeoff in mind, it is natural for
clients and curators to have different preferences regarding client
privacy. Some clients may value privacy more than others and thus will
tune their own privacy risk, while curators want to maximize their
model utility \emph{TorMentor is the first system to support anonymous
machine learning in a setting with heterogeneous user-controlled
privacy goals.}

\subsection{Training process}

Training in TorMentor (Algorithm~\ref{alg:training}) is performed in a
fashion similar to that of the parameter server~\cite{Li:2014}:
each client pulls the global model, locally computes a gradient step,
and applies the update to the global model. TorMentor uses the
differentially private SGD~\cite{Song:2013} method, which allows
clients to select their own privacy parameter $\varepsilon$ and batch
size $b$. We assume that clients understand how to properly define
these parameters and are aware of their implications on the
privacy-utility tradeoff and their privacy budgets~\cite{Dwork:2014}.

Since clients may fail or be removed from the system by the broker,
bulk synchronous computation in TorMentor may be infeasible. Instead,
as an alternative to the synchronous update model in federated
learning~\cite{McMahan:2017}, TorMentor also supports a total
asynchronous model~\cite{Hsieh:2017, Li:2014}, which enables
parallelization but allows clients to compute gradients on stale
versions of the global model, potentially compromising model
convergence. A lock-free approach to parallel SGD is feasible if
the the step size is tuned properly, and the corresponding global loss
function meets certain strong convexity guarantees~\cite{Recht:2011},
which we assume is true when using the total asynchronous model
in our brokered learning setting. This approach also negates the affect
of stragglers in a high latency environment (see 
Section~\ref{sec:eval}).


\begin{algorithm}[t]
  \KwData{Training data $x,y$; batch size $b$; privacy parameter
  $\varepsilon$}
  \KwResult{Returns a single gradient update on the model
  parameters} 
  \While{IsTraining} {
    Pull gradients $w_t$ from TorMentor\;
    Subsample $b$ points $(x_i, y_i) \in B_t$ from training data\;
    Draw noise $Z_t$ from Laplacian distribution\;
    Compute gradient step through differentially private SGD\;
    Push gradient to TorMentor
  }
  \caption{TorMentor differentially private SGD training algorithm.
  \label{alg:training}}
\end{algorithm}

Clients are free to leave the training process at any time. TorMentor
keeps a registry of the active clients, and checks that the minimum
number of clients condition is met at each gradient update. In
the case of clients leaving the system, TorMentor uses timeouts to
detect the clients who drop out of the system. Such clients do not
negatively impact the curator or other clients. As long as the required
minimum number of clients $k$ exists, the learning process will not
halt and no work will be wasted.

\subsection{Defending against inversion attacks}

\begin{figure}[t]
  \centering
  \includegraphics[width=.8\linewidth]{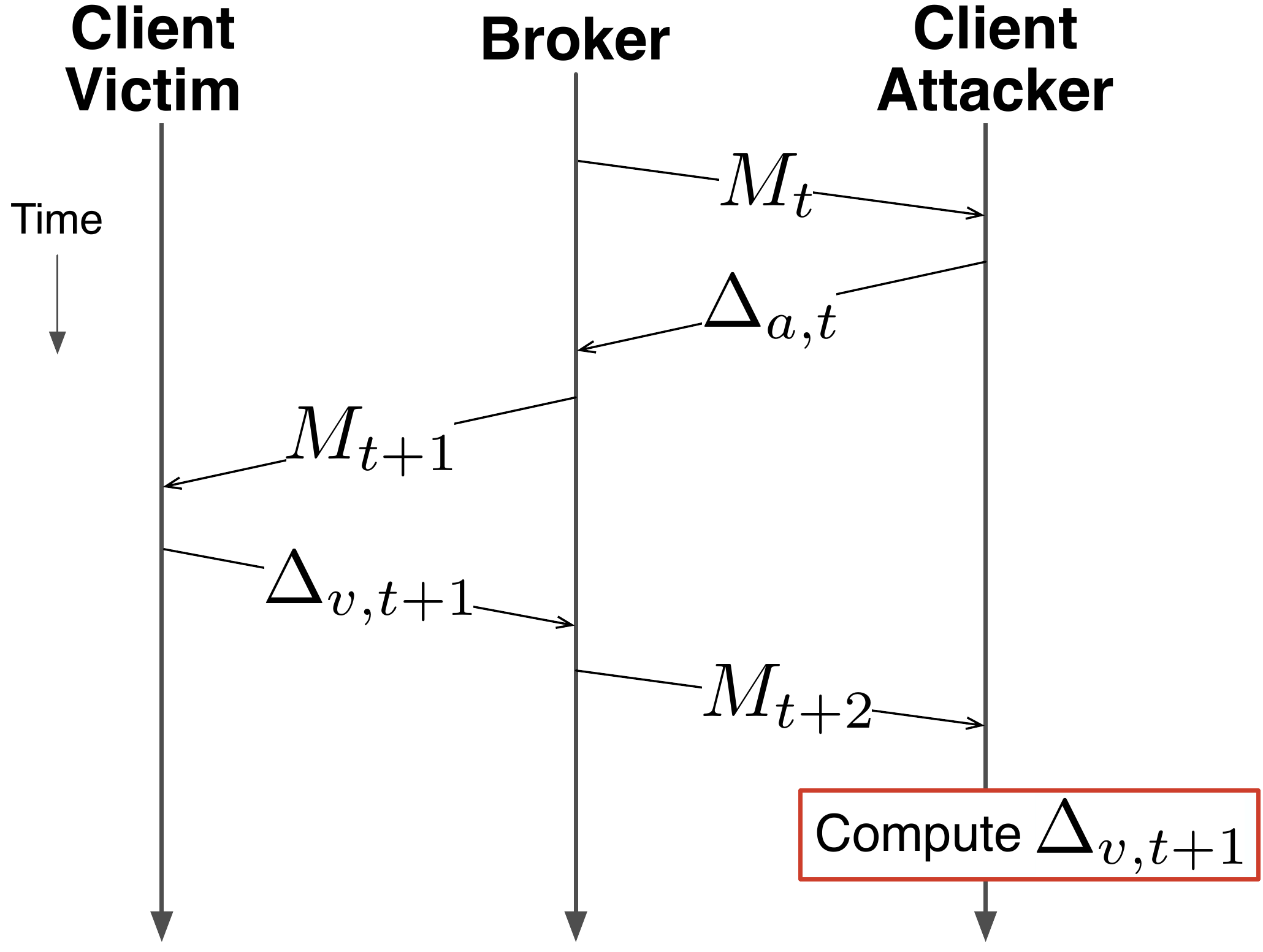}
  \caption{One iteration in an inversion attack in which an attacker
    observes the difference between $M_t$ and $M_{t+2}$, and infers
    this difference to be $\Delta_{V,t+1}$. After many iterations,
    the attacker can discover $M^*_V$, the optimal model trained on
    the victim's data.}
  \label{fig:inversion-timespace}
\end{figure}

Although a direct inversion attack in federated learning has not
been realized yet, we envision a novel potential attack in this
scenario. Figure~\ref{fig:inversion-timespace} shows the proposed ideal
situation for an attacker performing an inversion attack: a two client
TorMentor system, one of whom is the adversary.

In this scenario the victim $V$ and attacker $A$ alternate in sending
gradient updates to the broker. Since the global model parameters are
sent to the adversary at each iteration, it can ideally observe the
difference in the global model between iterations. As the attacker
knows their contribution to the global model at the previous
iteration, they are able to exactly compute the victim's update by
calculating:
\begin{align*}
  M_{t+2} &= M_t + \Delta_{v,t+1} + \Delta_{a,t} \\
  \Delta_{v,t+1} &= M_{t+1} - M_t - \Delta_{a,t}
\end{align*}
By saving and applying $\Delta_{v,t}$ at each iteration to a hidden,
shadow model, the adversary can compute an approximation to $M_V$,
the optimal model trained with only the victim's data, similar to a
model stealing attack~\cite{Tramer:2016}. The adversary can then
perform a model inversion attack~\cite{Fredrikson:2014,
Fredrikson:2015} and reconstruct the victim's training data elements
in $X_V$. In the case that the broker carries out the inversion attack,
the attack is even stronger: the broker can isolate all updates sent to
it through a single connection.

Differential privacy offers a natural defense against attacks from a
broker or another client by perturbing the victim's updates $\Delta_
{V,t}$ that are sent to the broker. An adversary will find it difficult
to recover $M_V$ and $X_V$ when the privacy parameter $\varepsilon$ is
closer to 0. An adversary could choose to send any vector as their
$\Delta_{a,t}$ update, which allows them to curate specific gradients
that elicit revealing responses from the victim~\cite{Hitaj:2017}.

In the case of attacks from other clients, the effectiveness of
the differentially private SGD protocol is also contingent on a setting
in which multiple clients are simultaneously performing updates. When
an adversarial client receives a new copy of the global model in
TorMentor, it has no mechanism to discover which clients contributed to
the model since the last update, making it difficult to derive
knowledge about specific clients in the system. 

Thus, TorMentor exposes a privacy parameter $k$ to clients, which
clients use to express the minimum number of clients that must be in
the system before training can begin. Our differentially private SGD
only begins when $n$ clients, each with a parameter $k \le n$ exist in
the system. Effectively, this means that for an adversarial client to
perform the ideal model inversion against a victim with parameter $k$
the adversary needs to create $k-1$ sybil clients.

\subsection{Defending against poisoning attacks}
\label{validator}

In adding the validator process, we propose an \emph{active parameter
server} alternative to the assumed passive parameter server in current
attacks~\cite{Hitaj:2017}. The parameter server validates each client's
contribution to the model health and penalizes updates from suspicious
connections. 

We develop a distributed RONI defense that uses sets of gradient
updates, instead of datasets, and independently evaluates the influence
that each gradient update has on the performance of a trusted global
model. Validation (Algorithm~\ref{alg:roni}) executes within the
parameter server in TorMentor and validates that iterations performed
by clients have a positive impact. Validation is parameterized by two
variables: the validation rate at which validations are performed, and
the RONI threshold~\cite{Barreno:2010} required before a client is
flagged.

To ensure that validations are performed in a fair manner, we benchmark
all clients against the same candidate model. The validator
intersperses validation iterations within the gradient updates requests
in distributed SGD. A validation round is triggered through a periodic 
Bernoulli test, with the probability parameterized by a validation
rate. During a validation round, the current model state is
snapshotted, and a copy of this model is sent to all active clients.
The clients' gradient responses are collected and the RONI value is
calculated and accumulated. 

In addition to the proof of work required to join the system, we
implement an adaptive proof of work mechanism to mitigate sybils in
poisoning attacks. A SHA-256 proof of work puzzle that must be solved
on each iteration before an update to the global model is accepted by
the broker. When a client's RONI score exceeds a defined negative
threshold, the broker increases the required trailing number of 0's by
one, increasing the difficulty of the puzzle for that client. The
difficulty of this puzzle is client- and iteration-specific, making it
more expensive for malicious clients to poison the model and decreasing
their influence relative to honest clients. 

\begin{algorithm}[t]
  \KwData{Stream of gradient updates from each client $i$, over $t$
  iterations $\Delta_{i,t}$}
  \KwResult{Reject a client if their updates oppose the defined
  learning objective}
  \While{IsTraining} {
    Draw Bernoulli value $v$\;
    \If{$v > \mathit{VALIDATION\mbox{\tt\string_}RATE}$} {
      Set current model $M_t$ to be snapshot model $M_s$\;
      Wait for client responses\;
    }
    \If{Client $c$ contacts TorMentor} {
      Send $M_s$ instead of $M_t$\;
      Save response $\Delta_{c,s}$
    }
    \If{All clients responded} {
       Find RONI $r_c$:
       $r_c = err(M_s, X_{val}) - err(M_s + \Delta_{c,s}, X_{val})$\; 
       $total_c = total_c + r_c$\;
      \If{$total_c > \mathit{THRESHOLD}$} {
        penalize $c$\;
      }
    }
  }
  \caption{RONI validation algorithm. \label{alg:roni}}
\end{algorithm}

When the rate of validation is increased, the broker discovers
poisoning clients more quickly, but with a performance overhead. When
the RONI threshold is increased, the broker is more likely to detect
adversaries, but the false positive rate of flagging honest nodes
increases as well.

An important design detail is that a validation request looks just
like a gradient update request. Therefore, adversaries cannot
trick the validator by appearing benign during validation rounds while
poisoning the true model. If the snapshot model $M_s$ is taken close
enough to the current model state $M_t$, it becomes more difficult for
adversaries to distinguish whether the updates they are receiving are
genuine gradient requests or not.

\subsection{Modular design}

To summarize, TorMentor includes several mechanisms to provide
various levels of client and curator privacy and security. These
include:
\begin{itemize}[label=$\star$]
    \item Using Tor to provide anonymous communication for all users 
    (clients and the curator).
    \item Using proof of work as a form of admission control.
    \item A validator process that uses RONI and adaptive proof of work
        to mitigate poisoning attacks and sybils.
    \item Differentially private SGD and minimum client enforcement to
        provide client privacy and to defend against inversion attacks.
\end{itemize}

Each of these components operates independently, and if brokered
learning is deployed in a setting where some of the described attacks
are out of scope, these components can be independently disabled. 

\section{TorMentor Implementation}
\label{sec:impl}

We implemented a TorMentor prototype in 600 LOC in Python 2.7 and
1,500 LOC in Go 1.8\footnote{
\url{https://github.com/DistributedML/TorML}}. All the communication
primitives are developed in Go, while the vector computation and ML
are in Python. We implement differentially-private 
SGD~\cite{Song:2013} in Numpy 1.12. For our noise function, we use a
multivariate isotropic Laplace distribution. As a performance
operation, we draw random samples from this distribution prior to
training by using \emph{emcee}, an MIT licensed Monte Carlo (MCMC)
ensemble sampler~\cite{mcmc:2013}.

In our evaluation we deploy the TorMentor curator and clients on
Azure by using bash scripts consisting of 371 LOC. These bootstrap VMs
with a TorMentor installation, launch clients, and orchestrate experiments.


\begin{figure*}[t]
	\includegraphics[width=\textwidth]{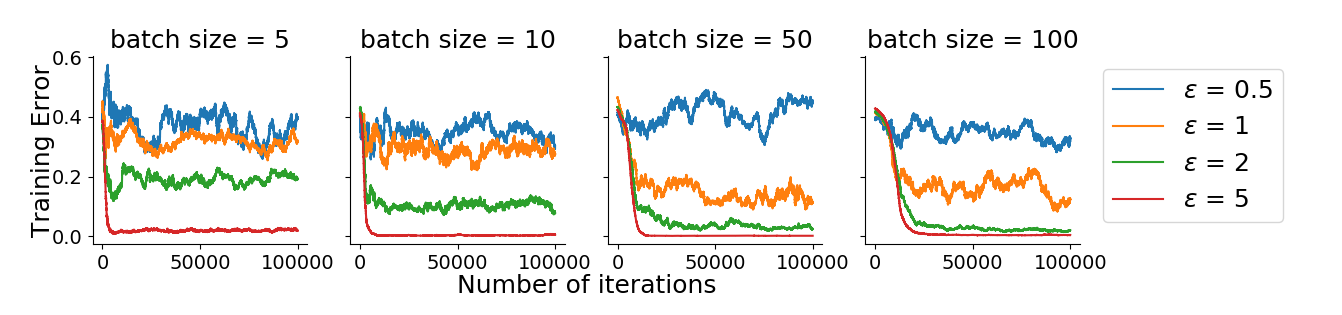}
	\caption{Effects of differential privacy and batch size on training
		loss over time. As $\varepsilon$ and $b$ decrease, model convergence
		is slower and may not be guaranteed.}
	\label{fig:bs50}
\end{figure*}

\section{Evaluation}
\label{sec:eval}

We evaluated our TorMentor design by carrying out several local
and wide-area experiments with the TorMentor prototype. Specifically,
we answer the following four research questions:

\begin{enumerate}

\item What are the effects of the privacy parameter $\varepsilon$ and
  batch size $b$ on model convergence?  (Section~\ref{eval:diffpriv})

\item What is TorMentor's overhead as compared to the baseline
  alternative? (Section~\ref{eval:overhead})
  
\item How effective are the privacy parameters $\varepsilon$ and
  minimum number of clients $k$ in defending against inversion
  attacks?  (Section~\ref{eval:inversion})
  
\item How effective is validation in defending against poisoning
  attacks, and what are the effects of its parameters?
  (Section~\ref{eval:poisoning})

\end{enumerate}

Next we describe the methodology behind our experiments and then
answer each of the questions above.

\subsection{Methodology}
\label{eval:method}

\textbf{Credit card dataset.} In our experiments we envision multiple
credit card companies collaborating to train a better model that
predicts defaults of credit card payments. However, the information in
the dataset is private, both to the credit card companies and to their
customers. In this context, any individual company can act as the
curator, the broker is a commercial trusted service provider, and
clients are the credit card companies with private datasets.

To evaluate this use-case we used a credit card
dataset~\cite{Yeh:2009} from the UCI machine learning
repository~\cite{Lichman:2013}. The dataset has 30,000 examples and
24 features. The features represent information about customers,
including their age, gender and education level, along with
information about the customer's payments over the last 6 months. The
dataset also contains information about whether or not the given
customer managed to pay their next credit card bill, which is used as
the prediction for the model.
 
Prior to performing the training, we normalized, permuted, and
partitioned the datasets into a 70\% training and 30\% testing shard.
For each experiment, the training set is further sub-sampled to create
a client dataset, and the testing shard is used as the
curator-provided validation set. Training error, the primary metric
used in evaluation, is calculated as the error when classifying the
entire 70\% training shard. In brokered learning, no single client
would have access to the entire training dataset, so this serves as a
hypothetical metric. \\

\noindent \textbf{Wide-area deployment on Azure.}  
We evaluated TorMentor at scale by deploying a geo-distributed set of
25 Azure VMs, each running in a separate data center, spanning 6
continents. Each VM was deployed using Azure's
default Ubuntu 16.06 resource allocation. Each VM was provisioned with
a single core Intel Xeon E5-2673 v3 2.40GHz CPU, and 4 GB of RAM.
Tor's default \texttt{stretch} distribution was installed on each client.
We deployed the broker at our home institution as a hidden service on
Tor. The median ping latency (without using Tor) from the client
VMs to the broker was 133.9ms with a standard deviation (SD) of 61.9ms.
With Tor, the median ping latency increased to 715.9ms with a SD of
181.8ms.

In our wide-area experiments we evenly distribute a varying number of
clients across the 25 VMs and measure the training error over
time. Each client joins the system with a bootstrapped sample of the
original training set (n = 21,000 and sampled with replacement), and
proceeds to participate in asynchronous model training.

\subsection{Model convergence}
\label{eval:diffpriv}

We evaluate the effect of the privacy parameter $\varepsilon$ and the
batch size $b$ when performing learning over TorMentor. 
Figure~\ref{fig:bs50} shows training error over time with a single
client performing differentially private SGD~\cite{Song:2013} to
train a logistic regression model using the entire training shard.

We found that models converge faster and more reliably when the batch
size is higher and when $\varepsilon$ is higher (less privacy). These
results are expected as they are confirmations of the utility-privacy
tradeoff. In settings with a low $\varepsilon$ (more privacy) and a low
batch size we observed that the effect of differential privacy is so
strong and the magnitude of the additive noise is so large that the
model itself does not converge, rendering the output of the model
useless. Based on these results, the experiments in the remainder of the
paper use a batch size of 10.

\subsection{Scalability and overhead}
\label{eval:overhead}

We also evaluated TorMentor's scalability by varying the number of
participating clients. We evaluate the overhead of Tor and the
wide-area in TorMentor by running TorMentor experiments with and
without Tor. All nodes were honest, held a subsample of the original
dataset, and performed asynchronous SGD.

\begin{figure}[t]
	\includegraphics[width=\linewidth]{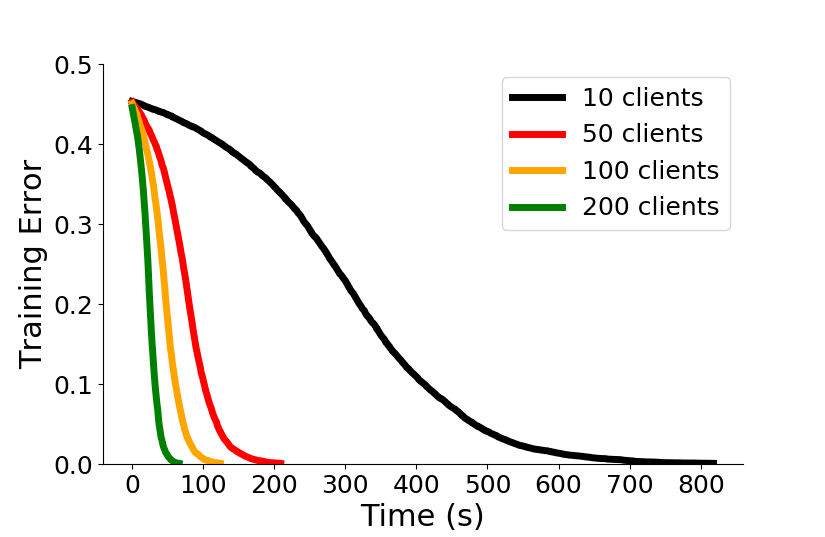}
	\caption{TorMentor model convergence in deployments with 10, 50,
          100, and 200 clients.
        }
	\label{fig:withtor}
\end{figure}

Figure~\ref{fig:withtor} shows that, when updating asynchronously, the
model convergences at a faster rate as we increase the number of
clients.

We also compared the convergence time on TorMentor with a baseline
WAN parameter server. For the WAN parameter server we used the same
clients deployment, but bypassed Tor, thereby sacrificing anonymity.

The results in Table~\ref{tab:latency} show that on average, the
overhead incurred from using Tor ranges from 5-10x. However, as the
number of clients increases, the training time in both deployments
drops, while the central deployment slows down.

\begin{table}[t]
\centering
\begin{tabular}{ c|cc }
 \hline
 \textbf{\# of Clients} & 
 \textbf{TorMentor}    & 
 \textbf{w/o Tor}  \\
 \hline
 10                    & 819 s & 210 s \\
 \hline
 50                    & 210 s  & 34 s \\
 \hline
 100                   & 135 s  & 18 s \\
 \hline
 200                   & 67 s   & 13 s \\
\end{tabular} 
\caption{Time to train the model with TorMentor, with and without Tor,
over a varying number of clients.
  \label{tab:latency}}
\end{table}

\begin{figure}[t]
	\includegraphics[width=\linewidth]{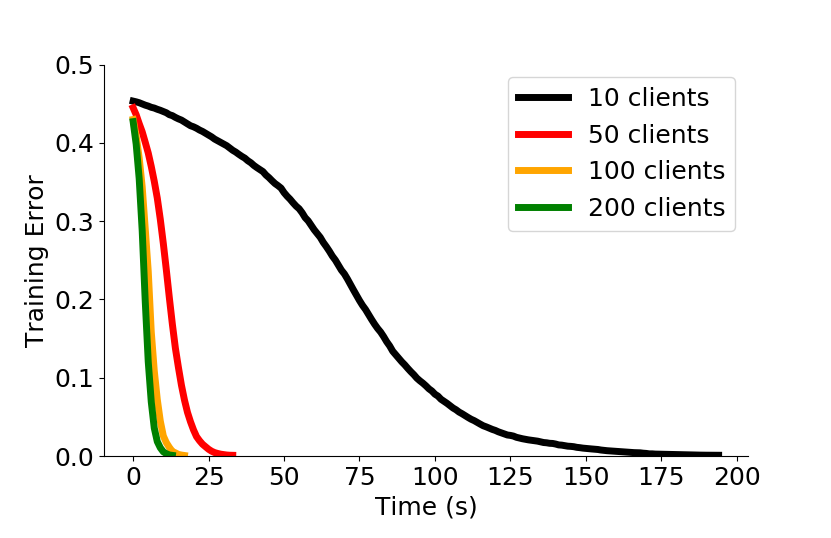}
	\caption{TorMentor without Tor model convergence in deployments with 10, 50,
          100, and 200 clients.
        %
        }
	\label{fig:without}
\end{figure}

\subsection{Inversion defenses evaluation}
\label{eval:inversion}

We first set up the inversion attack by carefully partitioning the
dataset. Of the 30,000 examples in the credit dataset, 9,000 (30\%)
examples were partitioned into $X_{test}$, a test dataset. The
remaining 21,000 examples were evenly partitioned across clients.
The victim's dataset $X_v$ was one of these partitions, with
all the $y_i$ prediction values flipped. This was done to
sufficiently separate the victim model from the globally trained
model\footnote{We originally attempted the inversion with one of the training data
shards as the victim dataset, but we found that even naively comparing
the final global model $M^*_g$ to the optimal victim model $M^*_v$
resulted in a low reconstruction error of 4.4\%. Thus, separating the
victim model in a way that makes it distinguishable is necessary.}. With
this victim dataset, a globally trained model achieved an error of
95.4\% when attempting to reconstruct the victim model, and predicting
on the test set.

With this setup we carried out the attack described in
Figure~\ref{fig:inversion-timespace}. Each attack was
executed for 4,000 gradient iterations, which was long enough for the
global model to reach convergence in the baseline case. We then
calculated the reconstruction error by comparing the resulting
inversion model to the true victim model, a model trained with only the
victim's data, by comparing predictions on the test set. That is, if
the inversion model and true victim model classify all test examples
identically, the reconstruction error is 0. The reconstruction error
would also serve as the error when an attacker uses outputs from
$\hat{M_v}$ to infer the training examples in $X_v$~\cite{Tramer:2016,
Fredrikson:2015}.

\begin{figure}[t]
	\includegraphics[width=\linewidth]{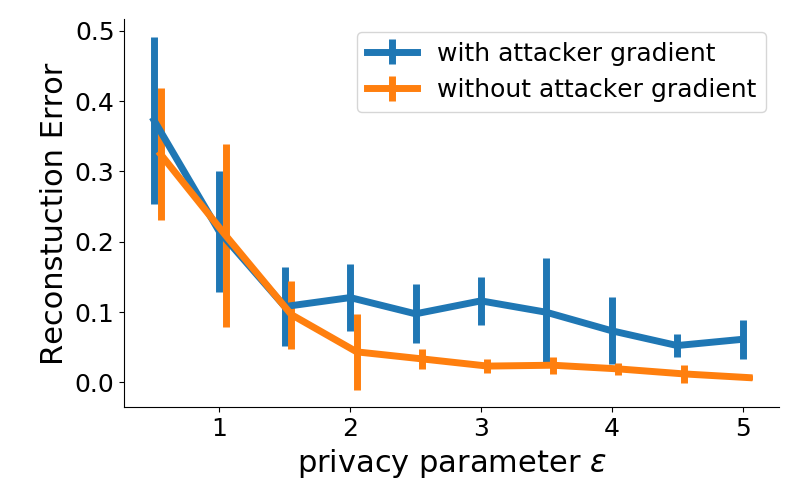}
	\caption{Model agreement between victim model and inverted estimate
	in the ideal setting (Fig.~\ref{fig:inversion-timespace}), with varying $\varepsilon$.}
	\label{fig:inversion-ideal}
\end{figure}

\begin{figure}[t]
	\includegraphics[width=\linewidth]{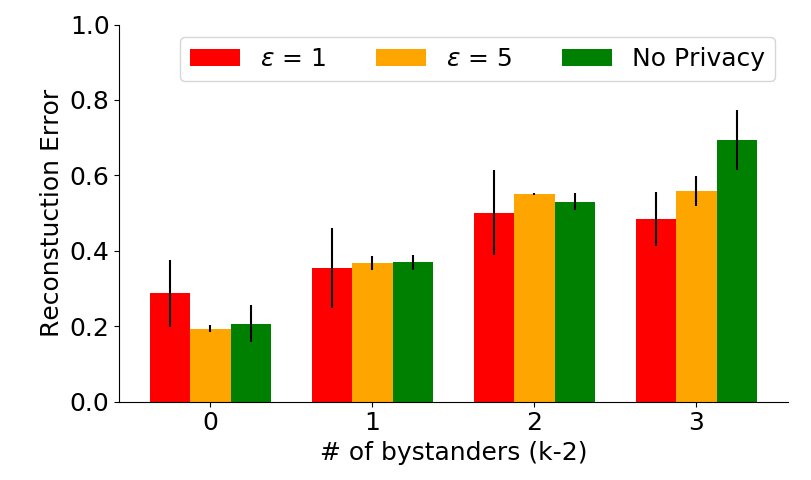}
	\caption{Reconstruction error between victim model and inverted
	model estimate, with varying privacy parameters: the number of
	bystanders and the privacy parameter $\varepsilon$.
          }
	\label{fig:inversion-bystander}
\end{figure}

Since the inversion attack is passively performed, it is defended by a
client carefully tuning the privacy parameters $\varepsilon$ and the
minimum number of clients $k$. We evaluate the effects of these
parameters in Figures~\ref{fig:inversion-ideal}
and~\ref{fig:inversion-bystander}.

Figure~\ref{fig:inversion-ideal} shows the effect of the privacy
parameter $\varepsilon$ on the reconstruction error, as $\varepsilon$
is varied from 0.5 to 5, plotting the median and standard deviation
over 5 executions.

In the baseline case the client and curator are alternating gradient
updates as in Figure~\ref{fig:inversion-timespace}, and there is no
differential privacy. As $\varepsilon$ decreases (increasing privacy),
the reconstruction error of the inversion attack increases. When
$\varepsilon = 1$, the reconstruction error is consistently 
above 10\%.

When the attacker sends a vector of zeros as their gradient update,
the inversion attack is most effective, as this completely isolates
the updates on the global model to those performed by the
victim. Figure~\ref{fig:inversion-ideal} shows the same experiment
performed when the attack contributes nothing to the global model. As
$\varepsilon$ increases beyond 2 (decreasing privacy), the attack
performed without sending any gradients consistently outperforms the
attack when performing gradient updates. This behavior, however, is
suspicious and a well designed validator would detect and blacklist
such an attacker. Therefore, this case is a worst case scenario as the
attacker must attempt to participate in the model training process.

Inversion attacks are made more difficult when randomness in the
ordering of gradient updates is increases. Two methods for increasing
this randomness include (1) adding random latencies at the broker, and 
(2) introducing bystanders: clients other than the attacker and victim.
In Figure~\ref{fig:inversion-bystander}, we evaluate both of these
methods by asynchronously training a model on TorMentor with one
victim, one attacker (using the same datasets as in 
Figure~\ref{fig:inversion-ideal}), and a varying number of bystanders.
When replying to a client response, we sample a random sleep duration
uniformly from 0-500ms at the server before returning a message.
All clients choose the same value for parameter $k$ and the actual
number of clients in the system is equal to $k$. Thus, in the framework
consisting of one victim and one attacker, the number of bystanders
equals $k-2$. 

Introducing even just one bystander ($k=3$) into the system increases
the reconstruction error during an inversion attack from about 20\% to
40\%. As $k$ grows, a model inversion attack becomes more difficult to
mount.

Figure~\ref{fig:inversion-bystander} also illustrates that differential
privacy defends client privacy when the number of bystanders is low.
When there are no bystanders in the system, decreasing the privacy
parameter $\varepsilon$ (more private) increases the reconstruction
error. The effects of a low $\varepsilon$ value in a model inversion
setting have a higher variance than in executions with higher
$\varepsilon$ values. Another mechanism that helps to mitigate
inversion attacks is the adaptive proof of work mechanism that counters
sybils (an attacker could spawn $k-1$ sybils as an alternative way to
isolate the victim).

\subsection{Poisoning defenses evaluation}
\label{eval:poisoning}

\begin{figure}[t]
	\includegraphics[width=\linewidth]{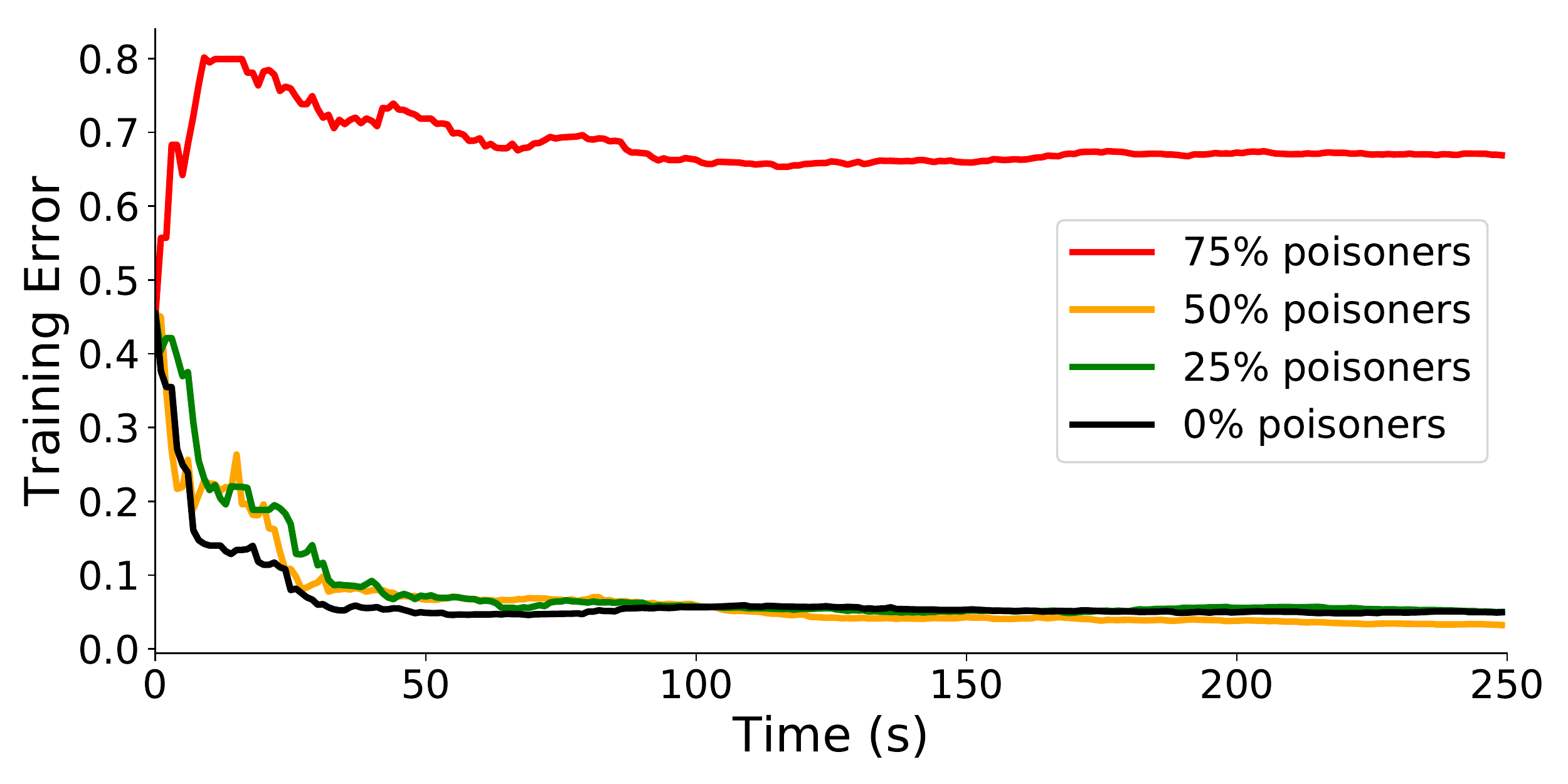}
	\caption{Model training loss over time, when attacked by a varying
	proportion of poisoners. RONI threshold is maintained at 2\%.}
	\label{fig:poison}
\end{figure}

We evaluate the effect of our proof of work on poisoning attacks. To
do this, we deployed TorMentor in an setting without differential
privacy or Tor in a total asynchronous setting with 8 clients. We
then varied the proportion of poisoners and the RONI threshold.
Figure~\ref{fig:poison} shows the training error for the first 250
seconds for a RONI threshold of 2\%, while varying the proportion of
poisoning attackers from 25\% to 75\%, with a validation rate of 0.1.

As the number of poisoners increases, different effects can be
observed. When the number of poisoners is low (below 25\%), the model
still converges, albeit at a slower rate than normal. With 50\%
poisoning, the model begins to move away from the optimum, but is
successfully defended by the validator, which increases the proof of
work required for all of the poisoners within 30 seconds. From this
point, the poisoners struggle to outpace the honest nodes, and the model
continues on a path to convergence. Lastly, when the proportion of
poisoners is 75\%, the increase in proof of work is too slow to
react; the model accuracy is greatly compromised within 20 seconds and
struggles to recover.

From this evaluation, we note that, if a poisoner was able to detect
this defense, and attempt to leave and rejoin the model, an optimal
proof of work admission puzzle should require enough time such that
this strategy becomes infeasible.

\begin{figure}[t]
	\includegraphics[width=\linewidth]{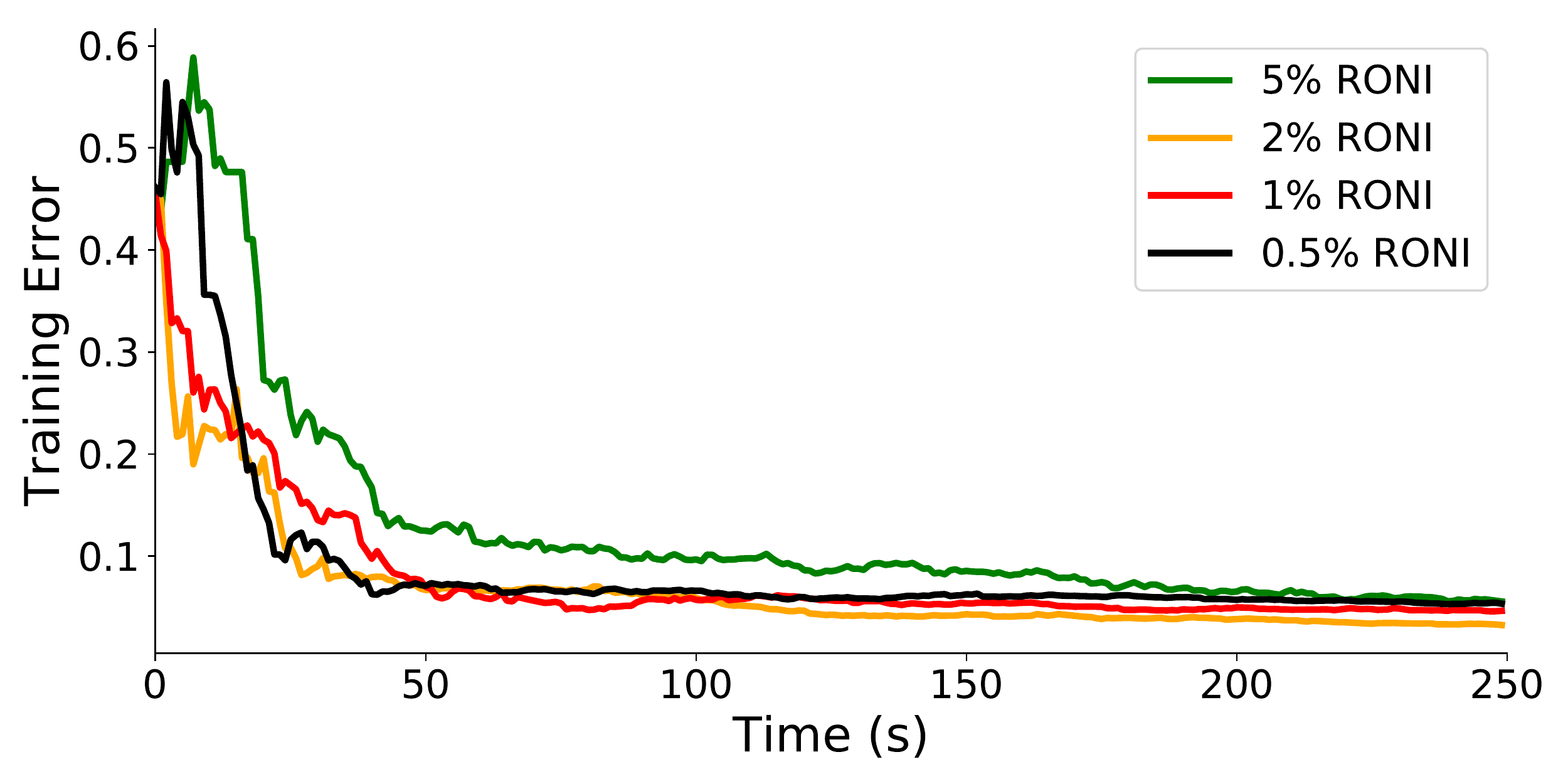}
	\caption{Model training loss over time, when attacked by 50\%
	poisoners. RONI threshold is varied from 0.5\% to 5\%. 
        }
	\label{fig:thresholds}
\end{figure}

Figure~\ref{fig:thresholds} shows the execution of model training with
50\% poisoning clients for different RONI validation thresholds. As
the threshold decreases, adversaries are removed from the system more
quickly, allowing the model to recover from the poisoning damage.

Setting the RONI threshold too low is also dangerous as it increases
the effect of false positives. In Figure~\ref{fig:thresholds}, we
observe that the model initially performs poorly, this is due to
incorrectly penalizing honest clients. The effect of a low
RONI is especially noticed in combination with differential privacy. To
confirm this, we performed two additional experiments in which the
validator had a RONI threshold of 0.5\% (the highest threshold from
Figure~\ref{fig:thresholds}), and a full set of honest clients with
differential privacy parameter $\varepsilon$ joined the model. When
$\varepsilon$ was set to 5, the model converged to an optimal point in
480 seconds. When $\varepsilon$ was set to 1, the validator flagged
all of the honest clients, and the model did not reach convergence.

The difference between model convergence, model divergence, and a
privacy violation all rely on a careful interplay between
$\varepsilon$, the minimum number of clients $k$, the RONI threshold,
the proof of work difficulty, and the anticipated attacks that
TorMentor expects to deter. Determining the optimal parameters for a
deployment depends on the anticipated workloads, data distribution, and
attack severity. Given the large scope of potential attacks and
attack scenarios in this space~\cite{Huang:2011}, we leave the
exploration of such parameter selection to future work.

\section{Discussion}
\label{sec:discuss}

\textbf{User incentives.}
Although TorMentor demonstrates that privacy-preserving, untrusted
collaborative ML is technically feasible, social feasibility remains an
open question~\cite{Stoica:2017}. That is, are there application
domains in which data providers are incentivized to contribute to
models from anonymous curators? And do these incentives depend on the
type of data, protections offered by a brokered learning setting, or
something else? Regarding curators, are there cases in which curators
would be comfortable using a model trained on data from anonymous
users? We believe that such application domains exist, partly because
of the widespread usage of federated learning~\cite{Frey:2017} and a
growing concern over data privacy~\cite{Fan:2015}.\\

\noindent \textbf{Usability of privacy parameters.}
Allowing clients and curators to define their own privacy parameters
$\varepsilon$ and $k$ allows for more expressive privacy policies, but
these parameters, and their implications, are difficult for users to
understand. Furthermore, privacy budgets, a fundamental element in
safely implementing and using differential privacy, are complex
and difficult to understand~\cite{Fan:2015}, as evidenced by Apple's
recent struggles in implementing such a system~\cite{Apple:2017}. \\

\noindent \textbf{Machine learning and Tor latency.}
Table~\ref{tab:latency} shows that Tor adds significant latency to the
machine learning process. On the one hand this (unpredictable) latency
can make it more difficult to mount an attack; for example, the
success of the inversion attack partly depends on predictable timing.
On the other hand, it would be desirable to shorten training time. 

At the moment Tor's latency is paid at each iteration, indicating that
methods with a lower iteration complexity would perform better. One
solution to this problem is to locally aggregate
gradients~\cite{Hsieh:2017, McMahan:2017, Bonawitz:2017} over many
iterations before sending them to the broker, trading off potential
model staleness for reduced communication costs. 

Outside of aggregating gradients, several iterative alternatives to SGD
exist, such as the Newton-Raphson method~\cite{Jennrich:1969} or other
quasi-Newton methods~\cite{Haelterman:2009}, which involve computing
the second-order Hessian. This provides convergence with a lower
iteration complexity, but with a higher computational cost per
iteration. A differentially-private version of the Newton-Raphson method
has also been developed~\cite{Ji:2014}.

TorMentor can be extended to generally support iterative ML update
methods. For models and learning objectives where Newton-Raphson is
applicable, we expect that Newton-Raphson will complete faster than SGD
when accounting for the overhead of Tor.\\

\noindent \textbf{Data-free gradient validation.}
While we demonstrated that our active validation approach defends against
attacks (e.g., Figure~\ref{fig:poison}), it relies on the curator, who
defines the learning objective, to provide the ground truth to
determine if an update is beneficial or not. This approach is not only
prone to bias but also opens up a new avenue for attack from the
curator; an adversarial curator could submit a junk validation set to
attack clients.

It is possible to mitigate these risks by using a data-free
solution. An open question is whether or not it is possible to achieve
the same effect without an explicit ground truth. We believe that the
use of a statistical outlier detection method~\cite{Hodge:2004} to
detect and remove anomalous gradient updates may bring the best of
both worlds. This would alleviate the need for and risk of a
curator-provided validation dataset, and this technique would also eliminate the
computational overhead of performing explicit validation rounds.
Another alternative would require the use of robust ML methods that are
known to handle poisoning attacks~\cite{Barreno:2010, 
Mozaffari-Kermani:2015}, but these methods are only applicable with
stronger assumptions about the curator who now must specify a ground
truth model.

\section{Conclusion}
\label{sec:conc}

We introduced a novel multi-party machine learning setting called
\textit{brokered learning}, in which data providers and model curators
do not trust one another and inter-operate through a third party
brokering service. All parties define their privacy requirements, and
the broker orchestrates the distributed machine learning process
while satisfying these requirements.
To demonstrate that this proposal is practical, we developed
TorMentor, a system for anonymous, privacy-preserving ML in the
brokered learning setting. TorMentor uses differentially private
model training methods to provide the strongest known defenses against
attacks in this setting~\cite{Huang:2011} and to support heterogeneous
privacy levels for data owners. We also developed and evaluated novel
ML attacks and defenses for the brokered learning setting.

Using a Tor hidden service as the broker to aggregate and validate client
gradient updates, TorMentor collaboratively trains a model across 200
geo-distributed clients, without ever directly accessing the raw data
or de-anonymizing any of the users. We define a realistic
threat model for brokered learning and show that in contrast to existing
solutions for distributed ML, such as Gaia~\cite{Hsieh:2017} and
federated learning~\cite{McMahan:2017}, TorMentor's defenses
successfully counter recently developed poisoning and inversion attacks
on ML. 



\subsection*{Acknowledgments}
We would like to thank Syed Mubashir Iqbal for his help in developing
bindings between Go and Python. We would also like to thank Margo
Seltzer, Mihir Nanavati, and Chris J.M. Yoon for their
feedback and comments. 
This research has been sponsored by the Natural Sciences and
Engineering Research Council of Canada (NSERC), 2014-04870.

{\footnotesize \bibliographystyle{acm}
\bibliography{main}}

\begin{appendices}
	\section*{Appendix A: SGD and differential privacy}

\noindent \textbf{Stochastic gradient descent (SGD).} In SGD at each iteration
$t$, the model parameters $w$ are updated as follows:

\begin{align*}
  w_{t+1} = w_t - \eta_t(\lambda w_t + \frac{1}{b} \sum_{(x_i, y_i)\in
  B_t} \nabla l(w_t, x_i, y_i)) &&&& (1)
\end{align*}
where $\eta_t$ represents a degrading learning rate, $\lambda$ is a
regularization parameter that prevents over-fitting, $B_t$ represents
a gradient batch of training data examples $(x_i, y_i)$ of size $b$ and
$\nabla l$ represents the gradient of the loss function. 

As the number of iterations increases, the effect of each gradient step
becomes smaller, indicating convergence to a global minimum of the loss
function. A typical heuristic involves running SGD for a fixed number
of iterations or halting when the magnitude of the gradient falls below
a threshold. When this occurs, model training is
complete and the parameters $w_t$ are returned as the optimal
model $w^*$.\\

\noindent \textbf{Distributed SGD.}
In parallelized ML training with a parameter server~\cite{Li:2014},
the global model parameters $w_g$ are partitioned and stored on a
parameter server. At each iteration, client machines, which house
horizontal partitions of the data, pull the global model parameters
$w_{g,t}$, compute and apply one or more iterations, and push their
update $\Delta_{i,t}$ back to the parameter server:

\begin{align*}
  \Delta_{i,t} = - \eta_t(\lambda w_{g,t} &+ \frac{1}{b} \sum_
  {(x_i, y_i)\in B_t} \nabla l(w_{g,t}, x_i, y_i)) && (2) \\
  w_{g,t+1} &= w_{g,t} + \sum_i \Delta_{i,t}
\end{align*}\\

\noindent \textbf{Differential privacy and SGD.}
\textit{$\varepsilon$-differential privacy} states that: given a
function $f$ and two neighboring datasets $D$ and $D'$ which differ in
only one example, the probability of the output prediction changes by
at most a multiplicative factor of $e^\varepsilon$. Formally, a
mechanism $f : D \to R$ is $\varepsilon$-differentially private for
any subset of outputs $S \subseteq R$ if
\[
Pr[f(D) \in S] \leq e^\varepsilon Pr[f(D') \in S].
\]

In differentially private SGD~\cite{Song:2013} the SGD update is
redefined to be the same as in Equation~(2), except with the addition
of noise:
\begin{align*}
  \Delta_{i,t} &= - \eta_t(\lambda w_{g,t} + \sum_{(x_i, y_i)\in B_t}
  \frac{\nabla l(w_{g,t}, x_i, y_i) + Z_t}{b}) && (3)
\end{align*}
where $Z_t$ is a noise vector drawn independently from a distribution:
\[
  p(z) \propto e^{(\alpha/2)||z||}
\]

\end{appendices}

\end{document}